\documentstyle[twocolumn,prb,eqsecnum,aps,graphicx,floats]{revtex}

\begin{document}
\draft
\twocolumn[\hsize\textwidth\columnwidth\hsize\csname@twocolumnfalse\endcsname

\title {Short range order in a steady state of irradiated 
Cu-Pd alloys: Comparison with fluctuations at thermal equilibrium}
\author{I. Tsatskis\cite{personal} and E. K. H. Salje}
\address{Department of Earth Sciences, University of Cambridge,
Downing Street, Cambridge CB2 3EQ, United Kingdom}
\maketitle
 
\begin{abstract}
The equilibrium short-range order (SRO) in Cu-Pd alloys is studied
theoretically. The evolution of the Fermi surface-related splitting
of the (110) diffuse intensity peak with changing temperature is
examined. The results are compared with experimental observations for
electron-irradiated samples in a steady state, for which the temperature
dependence of the splitting was previously found in the composition
range from 20 to 28 at.\% Pd. The equilibrium state is studied by
analysing available experimental and theoretical results and
using a recently proposed alpha-expansion theory of SRO which
is able to describe the temperature-dependent splitting. It is
found that the electronic-structure calculations in the framework
of the Korringa-Kohn-Rostoker coherent potential approximation
overestimate the experimental peak splitting. This discrepancy is
attributed to the shift of the intensity peaks with respect to the
positions of the corresponding reciprocal-space minima of the effective
interatomic interaction towards the (110) and equivalent positions.
Combined with an assumption about monotonicity of the temperature behaviour
of the splitting, such shift implies an increase of the splitting with
increasing temperature for all compositions considered in this study.
The alpha-expansion calculations seem to confirm this conclusion.
\end{abstract}


]

\section{Introduction}
\label{intro}

\begin{figure}
\begin{center}
\includegraphics[angle=0]{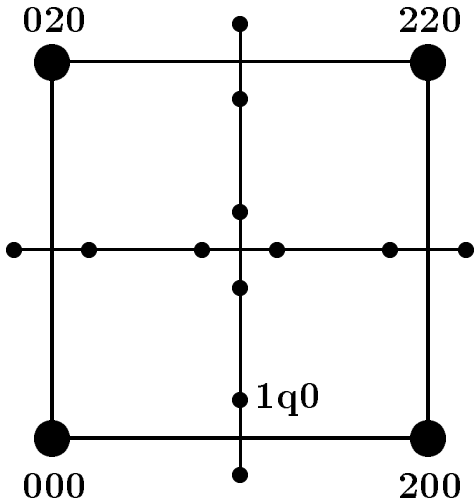}
\end{center}
\caption{Schematic reciprocal-space picture of scattering from 
disordered Cu-Pd alloys. Large dots represent the Bragg reflections. 
Small dots correspond to the split diffuse intensity peaks.}
\label{fig1}
\end{figure}

Almost a decade ago, Kulik {\em et al.}~\cite{kulik} published 
their experimental results on electron diffraction from irradiated 
Cu-Pd alloys. In that study samples with 20, 22, 24 and 28 
at.\% Pd were maintained by high-energy electron irradiation 
in a steady disordered state away from their thermal equilibrium 
state at temperatures between 200 and 400 K. At these temperatures 
and compositions equilibrium Cu-Pd alloys exhibit long-range order; 
the disordered state occurs only at much higher temperatures. 
In the equilibrium disordered state the intensity of diffuse 
scattering from Cu-Pd alloys with more than about 15 at.\% Pd is 
characterised by the fourfold splitting of intensity peaks 
located at the (100), (110) and equivalent positions in the reciprocal 
space.~\cite{kulik,ohshima1,watanabe,rodewald,saha,ohshima2,reichert2} 
The resulting diffuse intensity distribution has maxima  
at the $(1q0)$ and equivalent positions (Fig.~\ref{fig1}); the value 
of $q$ increases with increasing Pd concentration. This 
fine structure of diffuse scattering is caused by the atomic 
short-range order (SRO) and is a result 
of the indirect interaction of alloy atoms through conduction 
electrons in a situation when an alloy has reasonably well-defined 
Fermi surface with relatively flat areas.~\cite{krivoglaz,moss1}
In this case the corresponding minima of the effective pair 
interatomic interaction in the reciprocal space are also split, 
and their separation is related to the wavevector 
$2 {\bf k}_{F}$ which spans these flat areas of the Fermi surface.

Similar splitting of the diffuse intensity peaks was observed in 
the nonequilibrium steady disordered state under irradiation.~\cite{kulik} 
In addition to the expected concentration dependence of $q$, its 
variation with irradiation temperature was found. Even more curious 
was the qualitative change of the temperature dependence of the 
splitting with concentration: $q$ decreased with increasing temperature 
in the case of 20, 22 and 24 at.\% Pd, but this trend was reversed 
for the alloy with 28 at.\% Pd for which an increase of the peak 
separation with temperature was found. At the same time, there  
was an increase of the scattering intensity with 
increasing temperature for all four compositions, contrary to the 
case of alloys at equilibrium where the intensity decreases with 
increasing temperature. A qualitative explanation of the temperature 
dependence of the splitting was proposed as follows. Firstly, the 
behaviour of the equilibrium SRO diffuse intensity in the case of the exactly 
solvable one-dimensional Ising model with competing antiferromagnetic 
nearest- and next-nearest-neighbour interactions was studied. It 
turned out that the peak positions varied with 
temperature. This result is in contrast with the mean-field-related 
Krivoglaz-Clapp-Moss (KCM) treatment~\cite{krivoglaz,clapp} which 
predicts temperature-independent peak positions at the minima of 
the interaction. As the intensity increased with decreasing temperature, 
the peak positions shifted towards the wavevector of the corresponding 
ground state. A similar result was obtained earlier for the 
two-dimensional ANNNI model using the cluster variation method.~\cite{finel} 
It was concluded that the temperature dependence of the peak 
positions is a phenomenon which cannot be understood in the framework 
of mean-field theory. Secondly, the assumption was made that 
the behaviour of the diffuse intensity in irradiated Cu-Pd alloys was 
analogous to that of the equilibrium one-dimensional model. 
The only qualitative difference between the two cases was the opposite 
roles played by temperature. Based on this ``inverse temperature 
hypothesis'', the conclusion was drawn that one might expect to find 
the increase in $q$ with temperature for the Cu-Pd alloy system at 
equilibrium. Here it may be added that the reversal 
of the temperature behaviour of the peak splitting could be expected 
according to this hypothesis as concentration increases, from the 
increase with temperature for 20, 22 and 24 at.\% Pd to the decrease 
for 28 at.\% Pd. 

Last year, such an increase of the peak separation with temperature 
was observed at equilibrium by Reichert 
{\em et al.}~\cite{reichert} for the disordered Cu$_{3}$Au alloy. 
Moss and Reichert~\cite{moss2,reichert1} found the same behaviour by 
analysing the Monte Carlo simulation results of Roelofs 
{\em et al.}~\cite{roelofs} for the Cu-14.4 at.\% Al alloy. 
In the latter work the inverse Monte Carlo pair interactions 
were determined from the experimental diffuse intensity at a single 
temperature and subsequently used to generate the Monte Carlo intensities 
at other temperatures. The theory of the temperature dependence of the 
splitting was proposed by Tsatskis;~\cite{reichert1,tsatskis1} 
it identifies the wavevector dependence of the self-energy of the 
pair correlation function (PCF) as the origin of this effect. 
The self-energy $\Sigma({\bf k})$ and the interaction term 
$2 \beta V({\bf k})$ enter the expression for the SRO diffuse 
intensity on an equal footing (Eq.~(\ref{1a}) below). In the KCM 
approximation the fact that the self-energy is a function of 
${\bf k}$ is ignored. Apart from the observed increase in $q$ 
with temperature, the possibility of the opposite behaviour, i.e., 
the decrease of the peak separation as temperature increases, 
was predicted. This is exactly what should be expected under 
equilibrium conditions for the Cu-28 at.\% Pd alloy, if the inverse
temperature hypothesis is valid, although such temperature dependence 
was never seen experimentally. The possibility of the reversal
of the temperature dependence of the peak splitting with increasing 
concentration seems to be indicated also by the results of the 
Monte Carlo simulations of Ozoli\c{n}\u{s} {\em et al.}~\cite{ozolins} 
for the first-principles alloy Hamiltonian with pair and multiatom 
interactions (25 at.\% Pd) and the X-ray scattering 
measurements of Reichert {\em et al.}~\cite{reichert2} (29.8 at.\% Pd).  
In both cases no (or a very small) change of the splitting with 
changing temperature was found.

The idea of the present study is to gain further insight into 
the behaviour of Cu-Pd alloys under irradiation by studying theoretically 
the evolution of the diffuse peak splitting with changing temperature 
in these alloys at equilibrium. More exactly, the aim is to find out 
whether the splitting increases with temperature in the range from 20 
to 24 at.\% Pd as the inverse temperature hypothesis implies and whether 
this behaviour is reversed as concentration increases to 28 at.\% Pd. 
Starting from the experimental SRO diffuse intensity measured at a 
particular temperature, we first solve the inverse 
scattering problem and calculate the effective interaction which 
is assumed to be pairwise and temperature-independent. 
Than the direct problem is solved and the self-energy and 
diffuse intensity at different temperatures are calculated. 
The underlying theory of SRO is described in Sec.~\ref{theory}. 
Sec.~\ref{data} considers data for Cu-Pd alloys existing in the 
literature. Finally, the results are discussed in Sec.~\ref{results}. 

\section{Alpha-expansion theory of SRO}
\label{theory}

We start by describing the theory of SRO which leads to the 
temperature-dependent peak splitting and is used in Sec.~\ref{results} 
to relate the SRO diffuse intensities at different temperatures. 
This theory is based  on the alpha-expansion (AE) for the 
self-energy;~\cite{tsatskis1} the self-energy is the only unknown 
quantity in the otherwise formally exact expression for the SRO 
intensity. The AE is the expansion in powers of SRO parameters 
$\alpha_{lmn}$, hence the name. It was proposed as a generalisation 
of another approach to the calculation of the self-energy, the 
gamma-expansion method (GEM),~\cite{tokar1,tokar2,masanskii} to deal 
with distant interactions which are essential in the case of the Fermi
surface-related splitting. The complete set of the AE equations has 
the form
\begin{mathletters}
\label{1}
\begin{eqnarray}
I_{SRO}({\bf k}) & = & \frac{1}{ c(1-c) \left[ - \Sigma({\bf k}) 
+ 2 \beta V({\bf k}) \right] } \ , \label{1a} \\
\Sigma({\bf k}) & = & \Sigma_{000} + \sum_{lmn \neq 000} Z_{lmn} 
\Sigma_{lmn} \lambda_{lmn}({\bf k}) \ , \label{1b} \\
\Sigma_{lmn} & = & a \alpha_{lmn}^{2} 
+ b \alpha_{lmn}^{3} \ , \ \ \ lmn \neq 000 \ , \label{1c} \\
\alpha_{000} & = & \frac{1}{\Omega} \int d {\bf k} \, 
I_{SRO}({\bf k}) = 1 \ , \label{1d} \\
\alpha_{lmn} & = & \frac{1}{\Omega} \int d {\bf k} \, 
I_{SRO}({\bf k}) \, \lambda_{lmn}({\bf k}) \ . \label{1e}
\end{eqnarray}
\end{mathletters}
In Eqs.~(\ref{1}) ${\bf k}$ is the wavevector, $I_{SRO}({\bf k})$ 
is the SRO diffuse intensity in Laue units, $c$ is the concentration, 
$\Sigma({\bf k})$ is the self-energy of the PCF $G$ (the latter is 
defined by Eq.~(\ref{6}) below), $\beta=1/T$, $T$ is the temperature 
in energy units, and $V({\bf k})$ is the Fourier transform of the pair 
ordering potential 
\begin{equation}
V_{ij}=\frac{1}{2} \left( V^{AA}_{ij} + V^{BB}_{ij} \right) 
- V^{AB}_{ij} \ . \label{2}
\end{equation}
The potential $V^{\alpha \beta}_{ij}$ corresponds to the interaction 
between an atom of type $\alpha$ at site $i$ and an atom of type $\beta$ 
at site $j$. Further, $\alpha_{lmn}$, $\Sigma_{lmn}$, $Z_{lmn}$ and 
\begin{equation}
\lambda_{lmn}({\bf k}) = Z_{lmn}^{-1} \sum_{{\bf r} 
\in lmn} \exp (i {\bf k r}) \label{3}
\end{equation}
are the SRO parameter, matrix element of the self-energy, 
coordination number and shell function for the coordination 
shell $lmn$, respectively, while $\alpha_{000}$ and $\Sigma_{000}$ 
are the corresponding diagonal matrix elements. The summation in 
Eq.~(\ref{1b}) is performed over all coordination shells, whereas 
that in Eq.~(\ref{3}) is over the lattice vectors ${\bf r}$ belonging 
to the coordination shell $lmn$. The integration in Eqs.~(\ref{1d}) 
and  (\ref{1e}) is carried out over the Brillouin zone of volume 
$\Omega$. Coefficients $a$ and $b$ in Eq.~(\ref{1c}) are functions 
of concentration,
\begin{mathletters}
\label{4}
\begin{eqnarray}
a & = & \frac{(1-2c)^{2}}{2[c(1-c)]^{2}} \ , \label{4a} \\
b & = & \frac{[1-6c(1-c)]^{2}-3(1-2c)^{4}}{6[c(1-c)]^{3}} \ . \label{4b}
\end{eqnarray}
\end{mathletters}
The SRO parameters $\alpha$ are proportional to the corresponding 
matrix elements of the PCF $G$,
\begin{equation}
G^{AA}_{ij} = G^{BB}_{ij} = - G^{AB}_{ij} 
= c(1-c) \alpha_{ij} \ , \label{5}
\end{equation}
the definition of the PCF being
\begin{equation}
G^{\alpha \beta}_{ij} = \langle p^{\alpha}_{i} p^{\beta}_{j} 
\rangle - \langle p^{\alpha}_{i} \rangle \langle p^{\beta}_{j} 
\rangle \ , \label{6}
\end{equation}
where $p^{\alpha}_{i}$ is the occupation number,
\begin{equation}
p^{\alpha}_{i} = \left\{ \begin{array}{ll}
1, & \mbox{atom of type $\alpha$ at lattice site $i$} \ , \\
0, & \mbox{otherwise} \ , 
\end{array} \right. \label{7}
\end{equation}
and angular brackets denote statistical averaging.

The meaning of Eqs.~(\ref{1}) is as follows. The first of 
Eqs.~(\ref{1d}) and Eq.~(\ref{1e}) are the consequences of 
the fact that $\alpha_{ij}$ is the back Fourier transform of 
the intensity $I_{SRO}({\bf k})$. Eq.~(\ref{1b}) is the relation 
between the direct- and reciprocal-space representations 
of the self-energy. Eqs.~(\ref{1b}) and (\ref{1e}) are written in 
coordination shell notations. The second of Eqs.~(\ref{1d}) is the 
well-known sum rule~\cite{ducastelle} which reflects the property 
\begin{equation}
p^{\alpha}_{i} p^{\beta}_{i} = p^{\alpha}_{i} 
\delta^{\alpha \beta} \label{8}
\end{equation}
of the occupation numbers following from their definition~(\ref{7}). 
The less obvious Eq.~(\ref{1a}) is one of the possible 
forms of the Dyson equation~\cite{izyumov} which is satisfied by 
the PCF~(\ref{6}); this issue is discussed in considerable detail 
elsewhere.~\cite{tsatskis2}  The key equation is Eq.~(\ref{1c})
which closes the set of Eqs.~(\ref{1}) by expressing the 
off-diagonal part of the self-energy in terms of the SRO 
parameters. Its right-hand side is, in fact, two first non-zero 
terms of a series expansion of $\Sigma_{lmn}$ in powers of the SRO 
parameters. The latter are almost always sufficiently small, which 
justifies the expansion. These two terms were previously 
calculated~\cite{tokar2,masanskii} in the framework of 
the GEM using self-consistent renormalization of the bare propagator 
$(\beta V)^{-1}$ in the generating functional for correlation 
functions.~\cite{tokar1} The resulting expansion for the matrix elements of 
the self-energy was in powers of the matrix elements of the 
fully dressed propagator. This propagator is
the PCF~(\ref{6}), and its matrix elements are therefore 
proportional to the corresponding SRO parameters. Thus, Eqs.~(\ref{1}) 
form the set of self-consistent equations for the matrix elements of 
the self-energy (alternatively, $\Sigma_{000}$ and $\alpha_{lmn}$, 
$lmn \neq 000$, can be used as independent variables) and 
constitute the closed-form approximation for SRO. A particular 
AE approximation is defined by using Eq.~(\ref{1c}) for only a 
finite number of coordination shells and neglecting all other 
matrix elements of the self-energy. Another sequence of the AE 
approximations can be generated in the same way by taking into 
account only the lowest-order (quadratic) term in the AE
expansion for $\Sigma_{lmn}$ and ignoring the third-order 
contribution. For the rest of the paper both terms (as 
in Eq.~(\ref{1c})) will be used. The AE is expected to 
be at least as accurate as the GEM, and the latter was used 
successfully in dealing with both direct and inverse diffuse 
scattering problems,~\cite{tokar2,masanskii,reinhard} providing 
reliable results at almost all temperatures. The zero-order 
approximation of the AE is the well-known spherical model (SM) 
for correlations,~\cite{joyce} also known under the name of 
the Onsager cavity field theory.~\cite{onsager} In the SM the 
self-energy is diagonal, i.e., wavevector-independent; the 
single non-zero matrix element $\Sigma_{000}$ is a function of 
temperature and concentration and is determined from the sum 
rule~(\ref{1d}).

In order to use Eqs.~(\ref{1}) for calculating the evolution of the 
diffuse intensity with temperature and, in particular, the temperature 
dependence of the peak splitting, it is necessary to have information 
about the interaction $V$. It is assumed here that the interaction does not 
depend on temperature in the relevant temperature intervals; on the 
other hand, it is clearly concentration-dependent, so that a separate 
interaction set is needed for each alloy composition.
We start from the set of the experimental SRO parameters and calculate 
the AE interaction in the reciprocal space by solving the inverse 
diffuse-scattering problem.~\cite{tokar2,masanskii} This 
interaction can then be used for calculation of diffuse intensities at 
different temperatures and possibly, with much less confidence, 
at slightly different concentrations.

To solve the inverse problem, we rewrite Eq.~(\ref{1a}) as an 
expression for the interaction:
\begin{equation}
V_{AE}({\bf k}) = \frac{T}{2} \left[ 
\frac{I_{SRO}^{-1}({\bf k})}{c(1-c)} + \Sigma({\bf k}) 
\right] \ . \label{9}
\end{equation}
The SRO diffuse intensity here is recalculated from the set of the 
experimental SRO parameters:
\begin{equation}
I_{SRO}({\bf k}) = 1+ \sum_{lmn \neq 000} Z_{lmn} 
\alpha_{lmn} \lambda_{lmn}({\bf k}) \ . \label{10}
\end{equation}
In Eq.~(\ref{10}) the sum rule~(\ref{1d}) was used; otherwise, it is 
just the Fourier transformation written in coordination shell notations, 
similar to Eq.~(\ref{1b}). Substitution of Eqs.~(\ref{1b}) and (\ref{1c}) 
into Eq.~(\ref{9}) shows that the only quantity in the resulting 
expression for $V_{AE}({\bf k})$ which is needed to be expressed 
in terms of the SRO parameters (or, equivalently, the SRO 
intensity) is the diagonal part $\Sigma_{000}$ of the self-energy. 
The off-diagonal part of $\Sigma$ is already an explicit function 
of the SRO parameters (Eq.~(\ref{1c})). To find $\Sigma_{000}$, 
we integrate Eq.~(\ref{9}) over the Brillouin zone; 
this integration gives the diagonal direct-space matrix element 
of the integrand, as in Eq.~(\ref{1d}). The interaction $V$ is an 
off-diagonal matrix in the direct space because of the absence 
of the self-interaction. Therefore, after the integration the 
left-hand side of Eq.~(\ref{9}) is zero and, as a result,
\begin{mathletters}
\label{11}
\begin{eqnarray}
\Sigma_{000} & = & - \frac{\left\langle I_{SRO}^{-1} 
\right\rangle}{c(1-c)} \ , \label{11a} \\
\left\langle I_{SRO}^{-1} \right\rangle & = & \frac{1}{\Omega} 
\int d {\bf k} \, I_{SRO}^{-1}({\bf k}) \ . \label{11b} 
\end{eqnarray}
\end{mathletters}
Thus, Eq.~(\ref{9}) for the AE interaction can be written as
\begin{equation}
V_{AE}({\bf k}) = V_{SM}({\bf k}) + 
\frac{T}{2} \Sigma_{od}({\bf k}) \ , \label{12}
\end{equation}
where $\Sigma_{od}({\bf k})$ is the Fourier transform of 
the off-diagonal part of the self-energy defined by 
Eq.~(\ref{1c}), and 
\begin{equation}
V_{SM}({\bf k}) = \frac{T}{2c(1-c)} \left[ I_{SRO}^{-1}({\bf k}) - 
\langle I_{SRO}^{-1} \rangle \right] \label{13}
\end{equation}
is the interaction obtained in the framework of the SM, i.e., 
in the zero-order AE approximation in which the off-diagonal 
part of the self-energy is zero. Note that, when compared 
with the interaction resulting from the KCM expression for 
$I_{SRO}({\bf k})$,~\cite{krivoglaz,clapp}
\begin{equation}
V_{KCM}({\bf k}) = \frac{T}{2c(1-c)} \left[ 
I_{SRO}^{-1}({\bf k}) - 1 \right] \ , \label{14}
\end{equation}
the SM interaction differs, at given $c$ and $T$, only by the 
constant subtracted from the inverse intensity. Therefore, the 
off-diagonal direct-space interactions are identical in the KCM 
and SM approximations.~\cite{masanskii} However, the KCM formula 
violates the sum rule~(\ref{1d}), thus leading to the appearance 
of the unphysical self-interaction 
\begin{equation}
V^{KCM}_{000} = \frac{T}{2c(1-c)} \left[ 
\langle I_{SRO}^{-1} \rangle - 1 \right] \ , \label{15}
\end{equation}
while in the SM, according to Eq.~(\ref{13}), this matrix element 
is zero. Returning to Eq.~(\ref{12}), every term 
in its right-hand side is expressed at this stage in terms of 
experimental data, and $V_{AE}({\bf k})$ can be easily calculated.

\begin{figure}
\begin{center}
\includegraphics[angle=-90]{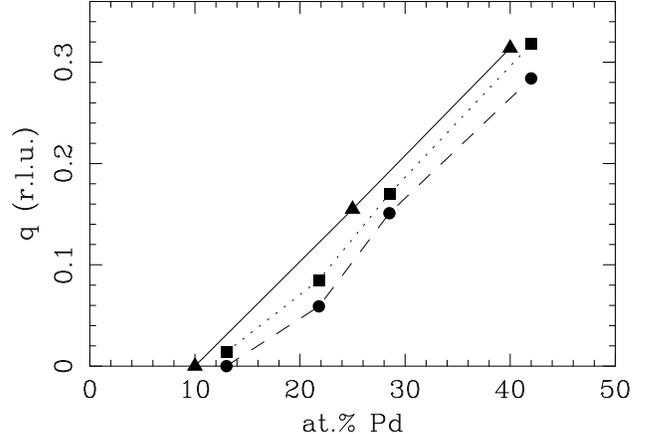}
\end{center}
\caption{Concentration dependence of the peak splitting $q$ as 
measured~\protect\cite{saha} in the X-ray diffraction experiment 
(circles, dashed line) and estimated~\protect\cite{saha} from the 
results of the KKR-CPA electronic-structure 
calculations~\protect\cite{gyorffy} (squares, dotted line). 
The data are taken from Table~1 in Ref.~\protect\onlinecite{saha}. 
The original KKR-CPA results~\protect\cite{gyorffy} (triangles, 
solid line) are also shown. Straight lines connecting symbols are
for the eye guidance only. Note that all the data were originally 
given in units of the distance between the (000) and (200) 
positions (equal to 2 r.l.u.) for the separation $m = \sqrt{2} \, q$ 
between the adjacent peaks.}
\label{fig2}
\end{figure}

\section{Available data}
\label{data}

We now consider previously published experimental and theoretical 
results for equilibrium Cu-Pd alloys in the discussed 
range of concentrations (20 to 30 at.\% Pd). These results are of 
two types. Firstly, the electron and X-ray diffraction data and 
the results of the Korringa-Kohn-Rostoker coherent potential 
approximation (KKR-CPA) electronic-structure calculations are available 
for the concentration dependence of the peak splitting. Secondly, for 
several alloy compositions large sets of the SRO parameters 
were determined by the Fourier inversion of the experimental SRO 
diffuse scattering intensities. The latter type of data 
is used as an input for the calculations of the kind 
described in Sec.~\ref{theory}.

\begin{figure}
\begin{center}
\includegraphics[angle=-90]{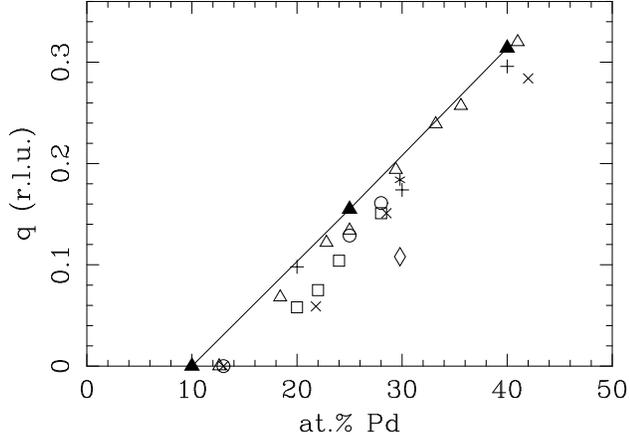}
\end{center}
\caption{The same as in Fig.~\ref{fig2}, but all available 
experimental data are presented 
(Ref.~\protect\onlinecite{kulik} - open squares, ca. 800 K; 
 Ref.~\protect\onlinecite{ohshima1} - open triangles, 773-893 K; 
 Ref.~\protect\onlinecite{watanabe} - open circles, ca. 700 K;
 Ref.~\protect\onlinecite{rodewald} - plusses, 1073 K;
 Ref.~\protect\onlinecite{saha} - crosses, 1023 K;
 Ref.~\protect\onlinecite{ohshima2} - asterisk, 773 K;
 Ref.~\protect\onlinecite{reichert2} - open diamond, ca. 700 K),
in comparison with the KKR-CPA results~\protect\cite{gyorffy} 
(filled triangles, solid line). The estimations made in 
Ref.~\protect\onlinecite{saha} are not shown.}
\label{fig3}
\end{figure}

The peak separation $q$ was measured at equilibrium 
for various concentrations and temperatures using 
electron~\cite{kulik,ohshima1,watanabe,rodewald} and 
X-ray~\cite{saha,ohshima2,reichert2} scattering. The splitting was 
observed for alloys with more than about 15 at.\% Pd, and it 
increased monotonically with increasing Pd content. Though very 
good agreement was noted by Gyorffy and Stocks~\cite{gyorffy} (GS)
between the electron-diffraction~\cite{ohshima1} and their 
KKR-CPA results, the calculated values of $q$ were systematically 
slightly higher than the experimental ones. The discrepancy 
became noticeably larger in more recent measurements. In particular, 
Saha {\em et al.}~\cite{saha} compared their X-ray results 
for several compositions with the estimations they made from the 
GS KKR-CPA calculations.~\cite{gyorffy} They found that the 
experimental splitting was smaller in all cases. The difference 
in $q$ ranged from 0.014 to 0.034 reciprocal lattice units 
(r.l.u.); 1 r.l.u. is the distance between the (000) and (100) 
positions. Their findings (Table~1 in Ref.~\onlinecite{saha}) 
are shown in Fig.~\ref{fig2}, together with the original GS results 
for different concentrations which were read off Fig.~2 in 
Ref.~\onlinecite{gyorffy}. Surprisingly, all the estimated values, 
presumably calculated by interpolating the results of 
Ref.~\onlinecite{gyorffy}, lie below the GS line;
the reason for this is not clear. As a result, the disagreement 
between the experimental and theoretical values of $q$ is even 
more pronounced than it was reported in Ref.~\onlinecite{saha}. 
Fig.~\ref{fig3} compares the KKR-CPA results with the collection 
of all experimental data for the splitting known to the authors. 
It is seen that all the experimental points are located below the 
GS line. We propose an explanation for this discrepancy which 
is given in Sec.~\ref{results}.

\begin{table}[b]
\caption{Data for three Cu-Pd alloys for which sets of the SRO 
parameters are available: $T$ is the annealing temperature, 
$N_{\alpha}$ the number of the SRO parameters in the set, 
$\alpha^{exp}_{000}$ and $q^{exp}$ the experimental values 
of $\alpha_{000}$ and $q$, respectively, $q^{rec}$ corresponds 
to the recalculated intensity (see text). The splitting $q$ is 
measured in r.l.u.}
\label{table1}
\begin{tabular}{cccccccc}
No. & at.\% Pd & Ref. & $T$, K & $N_{\alpha}$ & 
$\alpha_{000}^{exp}$ & $q^{exp}$ & $q^{rec}$ \\
\tableline
1 & 21.8 & \protect\onlinecite{saha}     & 1023 & 78 & 
1.018 & 0.059 & 0     \\
2 & 28.5 & \protect\onlinecite{saha}     & 1023 & 78 & 
1.014 & 0.151 & 0     \\
3 & 29.8 & \protect\onlinecite{ohshima2} &  773 & 72 & 
1.786 & 0.184 & 0.162 \\
\end{tabular}
\end{table}

\begin{figure}[t]
\begin{center}
\includegraphics[angle=-90]{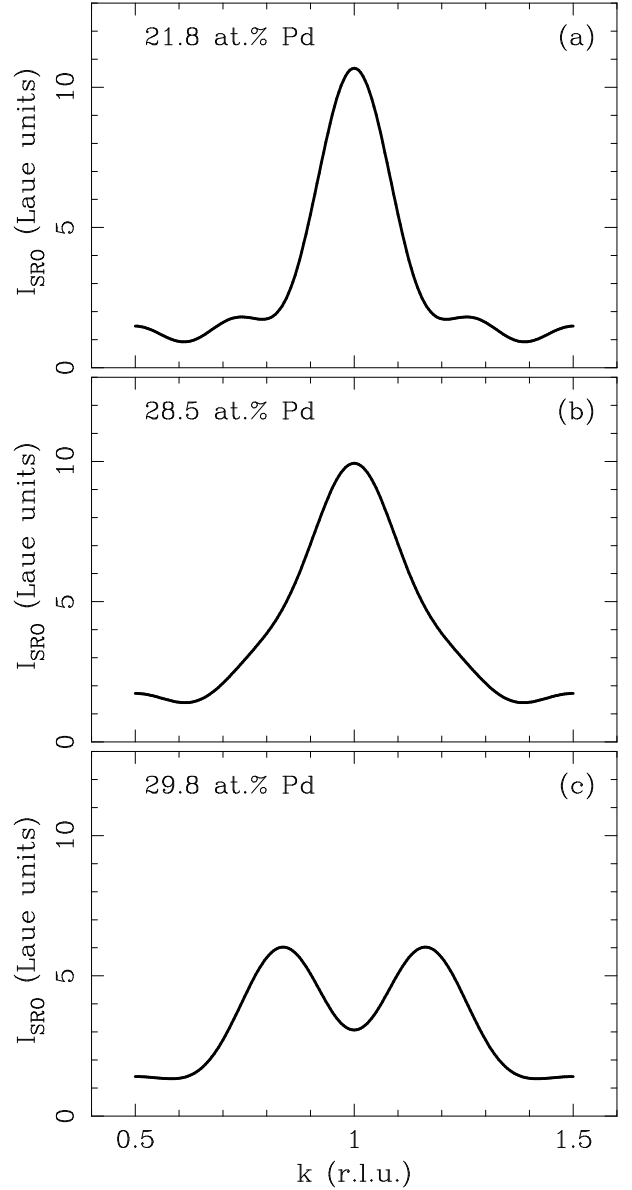}
\end{center}
\caption{Profiles of the recalculated SRO diffuse intensities for 
the alloys~1 (a), 2 (b), and 3 (c) along the (h10) line. Variable 
$k$ is the component of the wavevector ${\bf k}=(k,1,0)$. Note that 
there is no splitting of the (110) peak for the first two alloys.}
\label{fig4}
\end{figure}

\begin{figure}
\begin{center}
\includegraphics[angle=-90]{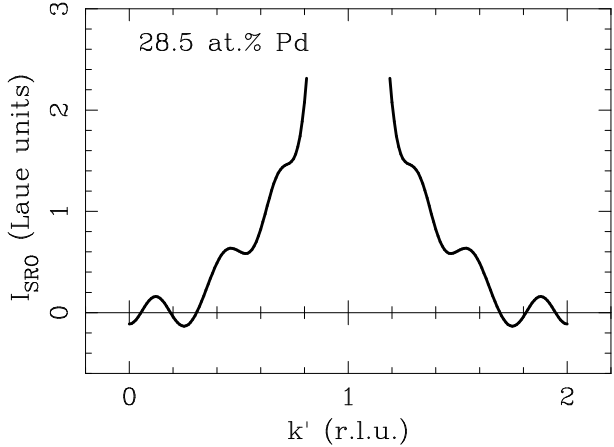}
\end{center}
\caption{Recalculated SRO intensity for the alloy~2 along the 
(h00) line showing ranges of negative values. Variable $k^{\prime}$ 
is the component of the wavevector ${\bf k}=(k^{\prime},0,0)$.}
\label{fig5}
\end{figure}

Sets of the SRO parameters were obtained in the considered 
concentration interval for 21.8, 28.5 (Ref.~\onlinecite{saha}) 
and 29.8 (Ref.~\onlinecite{ohshima2}) at.\% Pd in X-ray 
experiments. The samples were annealed at some temperature 
corresponding to the disordered phase and then quenched.
Hereafter these alloys will be referred to according to their 
numbers in Table~\ref{table1} which contains data used in the 
subsequent discussion. The splitting of the experimental 
(110) intensity peak was detected for all three compositions. 
The SRO parameters for large number of coordination shells were 
calculated by Fourier-transforming the SRO part of the measured 
diffuse intensity after having separated it from other intensity 
contributions. We recalculated the SRO diffuse intensities for these 
three alloys using tables of the SRO parameters given in 
Refs.~\onlinecite{saha} and \onlinecite{ohshima2} and the theoretical
value $\alpha_{000}=1$ instead of the experimental values. 
Such substitution leads to a simple shift of the intensity and 
does not change its shape. Surprisingly, no splitting of the 
(110) peak was found in the recalculated SRO intensities for the 
alloys~1 and 2 (Figs.~\ref{fig4}(a) and \ref{fig4}(b)); in 
the case of the alloy~2 this circumstance has already 
been noted elsewhere.~\cite{lu} In addition, negative values 
of the recalculated SRO intensity were found for the 
alloy~2 (Fig.~\ref{fig5}). The origin of all these inconsistencies 
is probably the insufficient accuracy and/or number of the 
calculated SRO parameters. Contrary to these two cases, the 
recalculated SRO intensity for the alloy~3 
shows the experimentally observed splitting (Fig.~\ref{fig4}(c)). 
However, the corresponding value of $q$ is noticeably smaller 
than the experimental result (Table~\ref{table1}). As for the first 
two alloys, we find that the splitting tends to decrease after 
the recalculation. The accuracy 
of the recalculated intensity seems to be better for the alloy~3 
as far as the magnitude of the splitting is concerned, though the 
deviation of the integrated intensity $\alpha_{000}$ from unity, 
which often serves as an accuracy criterion in diffuse-scattering 
experiments, is much larger than in two other cases 
(Table~\ref{table1}). 

\section{Results and discussion}
\label{results}

\begin{figure}
\begin{center}
\includegraphics[angle=-90]{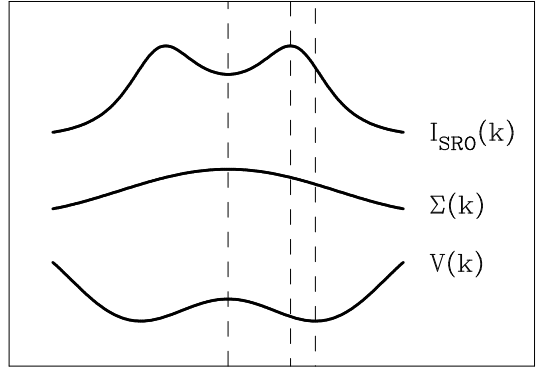}
\end{center}
\caption{Shift of the intensity peak position as a result of the
wavevector dependence of the self-energy. Behaviour of the SRO intensity,
self-energy and interatomic interaction along the (h10) line is shown 
schematically. The self-energy profile is as found for the three Cu-Pd 
alloys discussed in the text (see Fig.~\ref{fig8} below). Dashed lines 
indicate positions (left to right): (110), intensity peak, minimum of 
the interaction.}
\label{fig6}
\end{figure}

We assume that the discrepancy between the experimental and 
theoretical values of $q$ discussed in Sec.~\ref{data}
is the result of the shift of the intensity peak 
position with respect to the position of the corresponding minimum of 
the interatomic interaction.~\cite{tsatskis1} In other words, 
quantities which were measured and calculated were not the same.
Indeed, what GS actually calculated~\cite{gyorffy} 
using the KKR-CPA method were the Fermi surfaces and, in particular, 
the Fermi wavevectors ${\bf k}_{F}$ along the (110) direction
for different concentrations. These Fermi wavevectors were 
subsequently used to calculate the $2 {\bf k}_{F}$-related 
separation $m = \sqrt{2} \, q$ between the adjacent minima of 
the interaction $V({\bf k})$. Since the mean-field (KCM) 
description of correlations was chosen, the resulting separation 
between the intensity peaks was the same. However, it is generally 
different from the separation between the $V({\bf k})$ minima and depends 
on temperature because of the temperature-dependent shift of 
the intensity peak position.~\cite{tsatskis1} The shift itself 
is the consequence of the wavevector dependence of the self-energy 
(Fig.~\ref{fig6}). This can 
be easily seen from either Eq.~(\ref{1a}) (the direct problem) 
or Eqs.~(\ref{12}), (\ref{13}) (the inverse problem). Consider, 
e.g., Eq.~(\ref{1a}); the $I_{SRO}({\bf k})$ peak positions are 
determined by the condition ${\bf \nabla} I_{SRO} = 0$, which leads to 
\begin{equation}
2 \, {\bf \nabla} V = T \, {\bf \nabla} \Sigma \ , \label{16}
\end{equation}
while the positions of the $V({\bf k})$ minima are obtained from 
the equation ${\bf \nabla} V = 0$. It is clear from Eq.~(\ref{16}) 
that the extrema of $I_{SRO}({\bf k})$ away from the special points 
are, in general, different from those of $V({\bf k})$. On the other 
hand, if the approximate self-energy is ${\bf k}$-independent (as in 
the KCM or the SM approximations), then the two equations coincide 
and the intensity peak is not shifted. Fig.~\ref{fig3} shows 
that the KKR-CPA calculations overestimate the experimental 
peak splitting everywhere in the range of concentrations from 20 
to 30 at.\% Pd. In the framework of the suggestion about the shift 
of the $I_{SRO}({\bf k})$ peak being the origin of the disagreement 
between experiment and theory, this means that the intensity peaks 
are shifted towards the (110) position. 

Based on this assumption, it is now possible to predict 
the temperature behaviour of the splitting if another, 
sufficiently reasonable assumption is made. We assume that the 
temperature dependence of the splitting is always monotonic 
(a non-monotonic behaviour was never observed experimentally). 
If this assumption is correct, then the direction 
of the shift at a particular temperature value can be 
related to its temperature dependence. At high temperatures 
corrections to the KCM approximation are small, and the absolute 
value of the shift tends to zero, decreasing at least as $T^{-1}$ with 
increasing temperature.~\cite{tsatskis1} Therefore, in the case 
of the monotonic behaviour of the splitting the direction of the 
shift is the same at any temperature; the splitting increases 
with temperature, if the shift is towards the (110) position, and 
decreases otherwise. For Cu-Pd alloys this would mean that the 
splitting increases with temperature for all compositions in the 
considered range. This conclusion is in agreement with the one made 
on the basis of the inverse temperature hypothesis~\cite{kulik} 
discussed in Sec.~\ref{intro} for alloys with 20, 22 and 24 at.\% Pd, 
but it does not allow the change of the temperature behaviour 
predicted by this hypothesis for the Cu-28 at.\% Pd alloy.

\begin{figure}[t]
\begin{center}
\includegraphics[angle=-90]{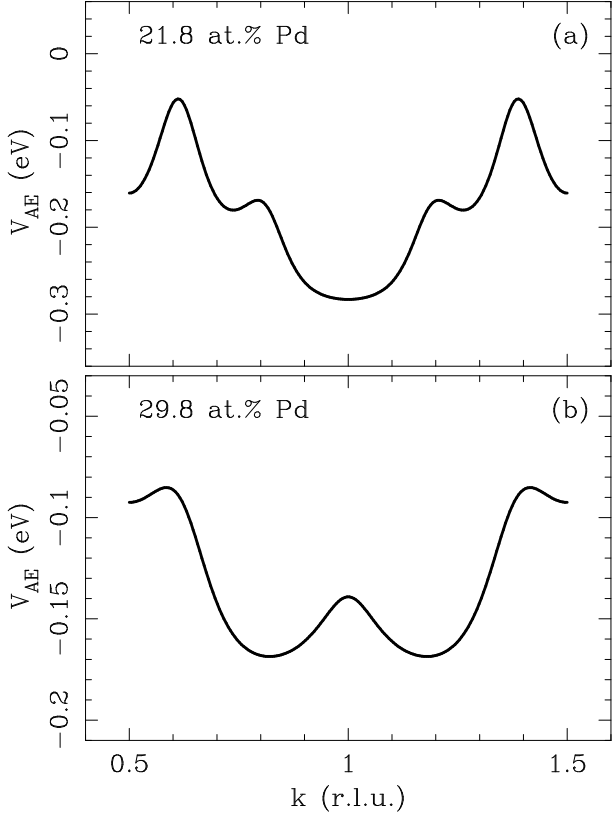}
\end{center}
\caption{AE effective pair interactions $V_{AE}({\bf k})$ for the 
alloys~1 (a) and 3 (b) along the (h10) line.}
\label{fig7}
\end{figure}

\begin{figure}[t]
\begin{center}
\includegraphics[angle=-90]{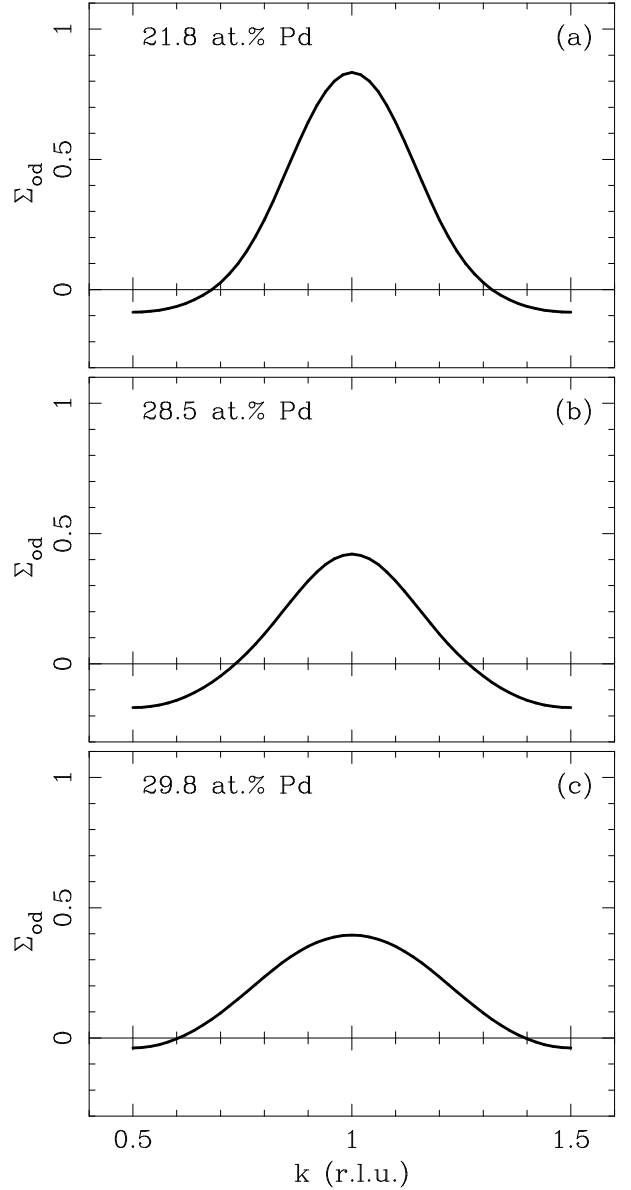}
\end{center}
\caption{Profiles of the off-diagonal part $\Sigma_{od}({\bf k})$ 
of the AE self-energy for the alloys~1 (a), 2 (b), and 3 (c) along 
the (h10) line. Maximal possible number $N_{\alpha}$ of coordination 
shells (Table~\ref{table1}) was used in each case.} 
\label{fig8}
\end{figure}

The next step is to check this prediction using the experimental 
data discussed in Sec.~\ref{data}. These data are quite limited, 
since sets of the SRO parameters are available only for three 
compositions. They are also not of sufficient accuracy for the 
reproduction of the fine structure of the (110) intensity peak. 
Only in one case, that of the 
alloy~3, the corresponding set is good enough (i.e., contains 
sufficient number of the reasonably accurate SRO parameters) to 
reproduce the experimentally observed splitting in the recalculated 
diffuse intensity (Fig.~\ref{fig4}). Even in this case, the 
recalculation changes noticeably the magnitude of the splitting
(see Table~\ref{table1}). It seems that in this particular situation 
of the split intensity peaks even larger sets of the more accurately 
determined SRO parameters are necessary. Nevertheless, the available 
sets can still be used to obtain information about the temperature 
dependence of the peak separation. The straightforward approach to 
this task described in Sec.~\ref{theory} is applicable only to the 
alloy~3. The solution of the inverse diffuse scattering problem 
given by Eqs.~(\ref{12}) and (\ref{13}) cannot be obtained for the 
alloy~2, because the recalculated diffuse intensity becomes negative 
(Fig.~\ref{fig5}); in this case the inverse intensity 
$I_{SRO}^{-1}({\bf k})$ and, therefore, the effective interatomic 
interaction $V_{AE}({\bf k})$ would contain unphysical singularities 
at those positions in the ${\bf k}$-space where the diffuse intensity 
vanishes. The inverse problem can be solved for 
the alloy~1, for which the recalculated intensity 
is always positive. However, in this case the resulting interaction
which follows the shape of the intensity does not have a split 
minimum at the (110) position. The profiles of $V_{AE}({\bf k})$ for 
the alloys~1 and 3 are shown in Fig.~\ref{fig7}. 

\widetext
\begin{figure}
\begin{center}
\includegraphics[angle=-90]{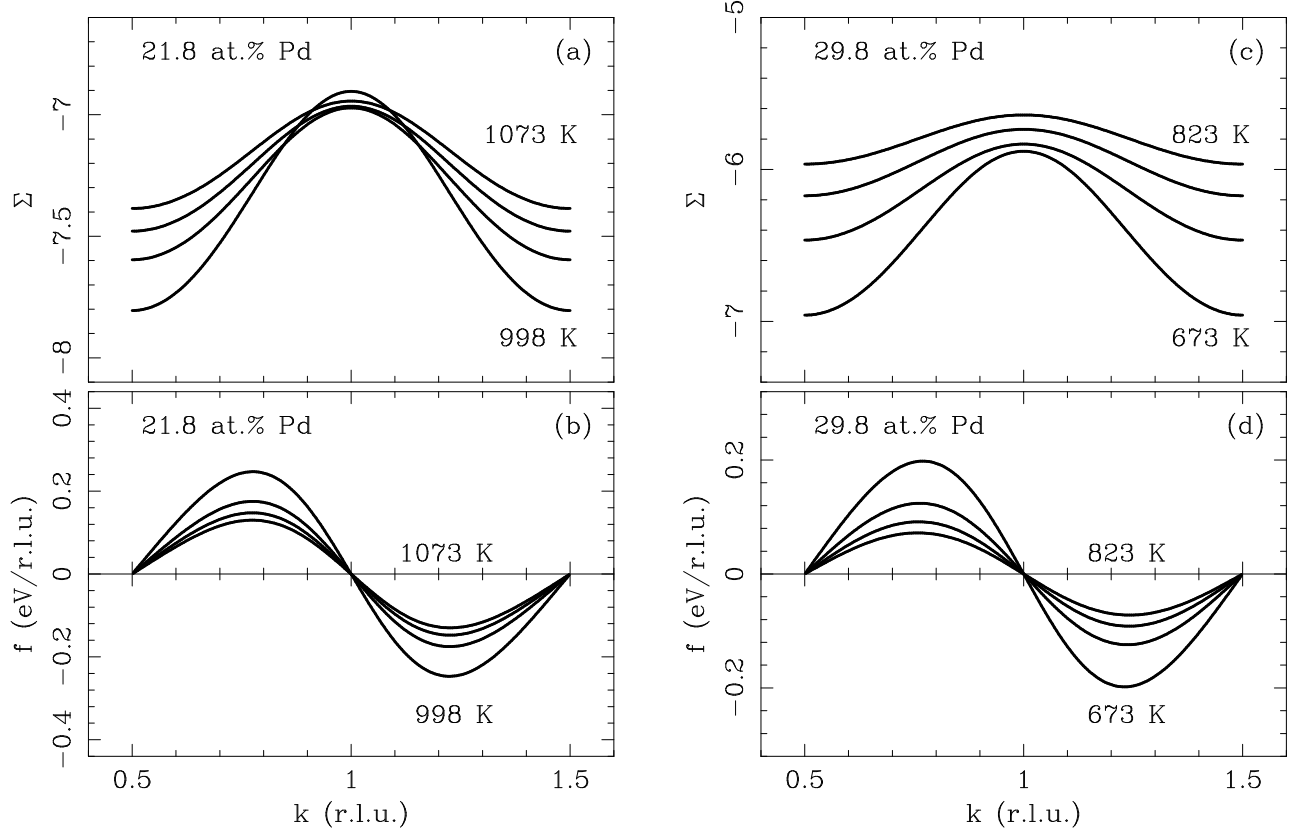}
\end{center}
\caption{AE self-energy $\Sigma(k)$ (a,c) and the related function 
$f(k)$ defined by Eq.~(\ref{17}) (b,d) for the alloys~1 (a,b) and 3 (c,d) 
along the (h10) line at several equidistant temperatures. Indicated are 
maximal and minimal temperatures; the temperature steps are 25 K (a,b) 
and 50 K (c,d). The KCM values for the self-energy (Eq.~(\ref{18})) are 
$\Sigma^{KCM}=-5.87$ (a) and $\Sigma^{KCM}=-4.78$ (c).}
\label{fig9}
\end{figure}
\narrowtext

The easiest quantity to calculate in the framework of the AE theory 
of SRO is the direction of the shift. This can be done for all 
three alloys, despite problems with the data for two of them 
as indicated before. According to Eq.~(\ref{16}), the direction 
of the shift is determined by the reciprocal-space behaviour of 
the self-energy, and the latter can be easily obtained by 
Fourier-transforming Eq.~(\ref{1c}), i.e., calculating 
$\Sigma_{od}({\bf k})$, and using the experimental values of 
the SRO parameters. It is expected that the off-diagonal part of 
the self-energy is much less sensitive to the accuracy of the 
$\{ \alpha_{lmn} \}$ set than the profile of the split intensity 
peak itself; there is no special reason for the self-energy to 
have any extrema away from the special points. Also, $\Sigma_{od}$ 
is of the second order in $\alpha_{lmn}$ (Eq.~(\ref{1c})) and 
therefore decreases in the direct space faster than the PCF, which 
means that the distant SRO parameters are less important for its 
calculation. The results of such calculation are presented 
in Fig.~\ref{fig8}. The convergence of the results with respect to 
the number of coordination shells included in the AE approximation 
improves rapidly with increasing concentration; to achieve very 
good convergence, about 40, 20 and 5 shells are necessary for the 
alloys~1, 2 and 3, respectively. In all three cases 
$\Sigma_{od}({\bf k})$ has a maximum at the (110) position, 
and from Eq.~(\ref{16}) it follows that the intensity peaks 
are shifted towards this position (see Fig.~\ref{fig6}). This result 
is in agreement with our interpretation of the discrepancy between 
the experimental and the KKR-CPA values of $q$. 

\begin{figure}
\begin{center}
\parbox[c][121mm][b]{86mm}{\hspace {1mm}}
\end{center}
\end{figure}

The actual value of the shift can be calculated only for the 
alloy~3, since for the other two alloys positions of neither peaks 
of the recalculated intensity nor minima of the AE interaction 
are available. The 10-shell AE approximation was used for this and 
all other calculations discussed in the rest of the paper. The 
result is 0.018 r.l.u.; it should be compared with the deviations 
of the experimental points from the GS line in Fig.~\ref{fig3}. 
The deviation for the alloy~3 calculated by the linear interpolation 
of the GS results is 0.022 r.l.u., which is very close to the result 
obtained in the AE calculation. If we assume that the AE shift is 
about the same for both the experimental and recalculated intensities, 
then the position of the $V_{AE}({\bf k})$ minimum for the former just 
falls on the GS line. The deviations of other experimental points are 
of the same order of magnitude. 

\begin{figure}
\begin{center}
\includegraphics[angle=-90]{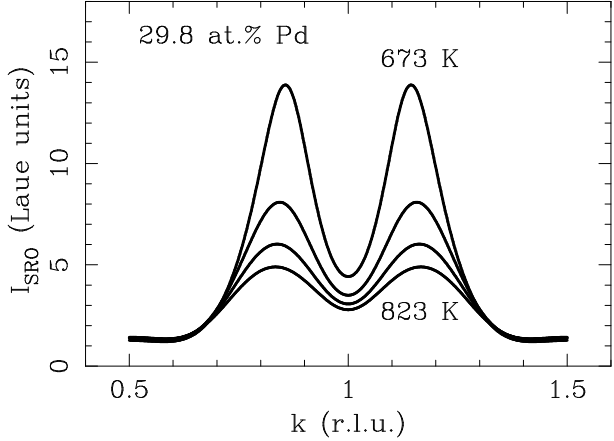}
\end{center}
\caption{AE intensity profiles for the alloy~3 along the (h10) line 
at several equidistant temperatures. The temperature range and step 
are as in Figs.~\ref{fig9}(c) and \ref{fig9}(d).}
\label{fig10}
\end{figure}

The change of the splitting with temperature can be analysed for 
the two cases (alloys~1 and 3) in which the inverse problem can be 
solved. This is done by calculating the self-energy as a 
function of temperature. Let us consider the profile 
of the self-energy along the (h10) line. Along this line the 
self-energy is a function of just one component $k$ of the 
wavevector ${\bf k}=(k,1,0)$. We define a function 
\begin{equation}
f(k) = T \, \frac{\partial \Sigma}{\partial k} \ , \label{17}
\end{equation}
the temperature dependence of which, according to Eq.~(\ref{16}), determines 
that of the splitting. The functions $\Sigma(k)$ and $f(k)$ at different 
temperatures for the two alloys are displayed in Fig.~\ref{fig9}. 
Accuracy checks show that the 10-shell approximation works 
very well for the alloy~3 and is still satisfactory 
(though noticeably worse) in the case of the alloy~1. 
Note that the AE results for the self-energy differ considerably 
from its KCM values; the KCM expression for the self-energy can be 
obtained, e.g., from the comparison of Eqs.~(\ref{9}) and (\ref{14}):
\begin{equation}
\Sigma^{KCM} = - \frac{1}{c(1-c)} \ . \label{18}
\end{equation}
In both cases the absolute value of $f(k)$ decreases with increasing 
temperature for any given value of $k$, which corresponds to the 
increase of the splitting with temperature. This result 
agrees with the conclusion based on the assumption of the monotonic 
temperature dependence of the splitting which was made earlier in 
this Section. 

\begin{figure}
\begin{center}
\includegraphics[angle=-90]{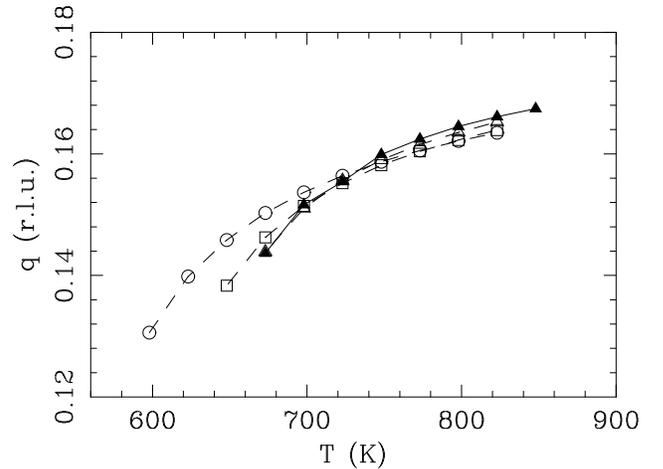}
\end{center}
\caption{Intensity peak position $q$ as function of temperature 
for the alloy~3 (filled triangles, solid line). Results for three 
other compositions (open symbols, dashed lines), 20 (circles), 
24 (squares) and 28 (triangles) at.\% Pd, calculated using the same 
AE interaction are also shown.}
\label{fig11}
\end{figure}

Finally, a quantitative calculation of the temperature dependence of 
the intensity peak position $q$ can be carried out, as before, for 
only one composition (alloy~3). The corresponding intensity profiles 
are presented in Fig.~\ref{fig10}, while Fig.~\ref{fig11} shows results 
for the function $q(T)$ for this composition, as well as for three 
other concentrations covering the interval which was considered in 
Ref.~\onlinecite{kulik}. In all calculations the same AE interaction 
(namely, that obtained by solving the inverse problem for the alloy~3;
see Fig.~\ref{fig7}(b)) was used. The aim was to find out whether the 
variation of composition for the same interaction would lead to any 
particular change of the function $q(T)$. No such change takes place, 
as it is seen from Fig.~\ref{fig11}; the splitting increases 
monotonically with temperature for all four alloy concentrations.

In summary, we studied theoretically the temperature dependence of 
the Fermi surface-induced splitting of the (110) SRO diffuse intensity 
peak for Cu-Pd alloys under equilibrium conditions. The comparison was 
made with experimental observations for these alloys in a steady state 
under irradiation. The validity of the inverse temperature hypothesis 
proposed previously to relate the two regimes was examined. At 
equilibrium this hypothesis predicts the qualitative change of 
temperature behaviour near the Cu$_{3}$Pd composition, namely, the 
increase of the splitting with increasing temperature in the composition 
interval 20 to 24 at.\% Pd and its decrease with temperature as 
the concentration of Pd increases to 28 at.\%. Comparing available 
electron and X-ray scattering data with the results of the KKR-CPA 
electronic-structure calculations, we found that the theoretical 
approach overestimated the experimental splitting. This disagreement 
was interpreted as the result of the shift of the diffuse intensity 
peaks with respect to the positions of the corresponding minima 
of the effective pair interatomic interaction towards the (110) 
position. An additional assumption about monotonicity of the temperature 
dependence of the splitting led to a connection between the direction and 
temperature behaviour of the peak shift. Under this assumption the shift 
towards the (110) position is equivalent to the increase of the splitting 
with increasing temperature. For Cu-Pd alloys this means that the 
splitting increases with temperature for all concentrations in the 
considered compositional range. This conclusion seems to be confirmed by 
the AE calculations, which are, however, based on limited experimental data. 
It agrees with the prediction of the inverse temperature hypothesis for 
lower Pd concentrations (20 to 24 at.\%) but, contrary to this prediction, 
does not allow any reversal of the temperature behaviour with increasing 
concentration of Pd. It also contradicts the results of recent computer 
simulations~\cite{ozolins} and X-ray measurements~\cite{reichert2} for 
higher Pd content (25 and 29.8 at.\%, respectively), according to which 
the splitting is (almost) temperature-independent. These results are 
consistent, on the other hand, with the reversal scenario. Among possible 
reasons for this disagreement are (i) limited validity of the inverse 
temperature hypothesis, (ii) insufficient accuracy and/or size of the 
available sets of the experimental SRO parameters, (iii) pair character 
of the interatomic interactions used in the AE theory of SRO and  (iv) 
approximate character of the AE calculations. Further direct 
measurements of the splitting as a function of temperature (as in 
Ref.~\onlinecite{reichert2}) at the discussed range of compositions 
are necessary to clarify the situation.

The authors would like to thank H. Reichert and collaborators for communicating
their experimental results (Ref.~\onlinecite{reichert2}) prior to publication.


\begin{references}

\bibitem[\dag]{personal}
Former name: I. V. Masanskii

\bibitem{kulik}
J. Kulik, D. Gratias, and D. de Fontaine, Phys. Rev. B 
{\bf 40}, 8607 (1989).

\bibitem {ohshima1}
K. Ohshima and D. Watanabe, Acta Cryst. A {\bf 29}, 520 (1973).

\bibitem{watanabe}
D. Watanabe, J. Phys. Soc. Jpn. {\bf 14}, 436 (1959).

\bibitem{rodewald}
M. Rodewald, K. Rodewald, P. De Meulenaere, and G. Van Tendeloo, 
Phys. Rev. B {\bf 55}, 14173 (1997).

\bibitem{saha}
D.K. Saha, K. Koga, and K. Ohshima, J. Phys. Condens. Matter
{\bf 4}, 10093 (1992).

\bibitem{ohshima2}
K. Ohshima, D. Watanabe, and J. Harada, Acta Cryst. 
A {\bf 32}, 883 (1976).

\bibitem{reichert2}
H. Reichert, H.H. Hung, V. Jahns, K.S. Liang, D. Zehner, and H. Dosch
(to be published).

\bibitem{krivoglaz}
M.A. Krivoglaz, {\em Theory of X-Ray and Thermal Neutron 
Scattering by Real Crystals} (Plenum, New York, 1969); 
{\em Diffuse Scattering of X-Rays and Neutrons by 
Fluctuations} (Springer, Berlin, 1996).

\bibitem{moss1}
S.C. Moss, Phys. Rev. Lett. {\bf 22}, 1108 (1969);
S.C. Moss and R.H. Walker, J. Appl. Crystallogr. {\bf 8}, 96 (1974);

\bibitem{clapp}
P.C. Clapp and S.C. Moss, Phys. Rev. {\bf 142}, 418 (1966);
{\bf 171}, 754 (1968).

\bibitem{finel}
A. Finel and D. de Fontaine, J. Statist. Phys. {\bf 43}, 645 (1986).

\bibitem{reichert}
H. Reichert, S.C. Moss, and K.S. Liang, Phys. Rev. Lett. 
{\bf 77}, 4382 (1996).

\bibitem{moss2}
S.C. Moss and H. Reichert (private communication). 

\bibitem{reichert1}
H.~Reichert, I. Tsatskis, and S.C. Moss, in {\em Proceedings of 
the Joint NSF/CNRS Workshop on Alloy Theory, Mont Sainte Odile 
Monastery, Strasbourg, France, October 11-15, 1996}, 
Comput. Mater. Sci. {\bf 8}, 46 (1997).

\bibitem{roelofs}
H. Roelofs, B. Sch\"{o}nfeld, G. Kostorz, W. B\"{u}hrer, 
J.L. Robertson, P. Zschack, and G.E. Ice, Scripta Mat. 
{\bf 34}, 1393 (1996).

\bibitem{tsatskis1}
I. Tsatskis, preprint cond-mat/9801089 (to be published).

\bibitem{ozolins}
V. Ozoli\c{n}\u{s}, C. Wolverton, and A. Zunger, Phys. Rev. Lett. 
{\bf 79}, 955 (1997).

\bibitem{tokar1}
V.I. Tokar, Phys. Lett. A {\bf 110}, 453 (1985).

\bibitem{tokar2}
V.I. Tokar, I.V. Masanskii, and T.A. Grishchenko,
J. Phys.: Condens. Matter {\bf 2}, 10199 (1990).

\bibitem{masanskii}
I.V. Masanskii, V.I. Tokar, and T.A. Grishchenko, 
Phys. Rev. B {\bf 44}, 4647 (1991).

\bibitem{ducastelle}
E.g., F. Ducastelle, {\em Order and Phase Stability in Alloys} 
(North-Holland, Amsterdam, 1991).

\bibitem{izyumov}
E.g., Yu.A. Izyumov and Yu.N. Skryabin, {\em Statistical 
Mechanics of Magnetically Ordered Systems} (Consultants Bureau, 
New York and London, 1988).

\bibitem{tsatskis2}
I. Tsatskis, in {\em Local Structure from Diffraction}, Fundamental 
Materials Science Series, edited by M.F.~Thorpe and S.J.L.~Billinge 
(Plenum Press, New York, 1998, in press).

\bibitem{reinhard}
L. Reinhard and S.C. Moss, Ultramicroscopy {\bf 52}, 223 (1993);
M. Borici-Kuqo and R. Monnier, Ref.~\onlinecite{reichert1}, p.~16;
D. Le Bolloc'h, T. Cren, R. Caudron and A. Finel, 
Ref.~\onlinecite{reichert1}, p.~24. 

\bibitem{joyce}
G.S. Joyce, in {\em Phase Transitions and Critical Phenomena}, 
Vol.~2, eds. C.~Domb and M.S.~Green (Academic Press, New York, 
1972); R. Brout, {\em Phase Transitions} (Benjamin, New York, 1965).

\bibitem{onsager}
L. Onsager, J. Am. Chem. Soc. {\bf 58}, 1468 (1936); J.B. Staunton 
and B.L. Gyorffy, Phys. Rev. Lett. {\bf 69}, 371 (1992).

\bibitem{gyorffy}
B.L. Gyorffy and G.M. Stocks, Phys. Rev. Lett. 
{\bf 50}, 374 (1983).

\bibitem{lu}
Z.W. Lu, D.B. Laks, S.-H. Wei, and A. Zunger, Phys. Rev. B 
{\bf 50}, 6642 (1994).

\end{references}
\end{document}